\documentclass[conference]{IEEEtran}
\IEEEoverridecommandlockouts
\usepackage{cite}
\usepackage{amsmath,amssymb,amsfonts}

\usepackage{mathtools}

\usepackage{cuted}
\usepackage{graphicx}
\usepackage{textcomp}
\usepackage{xcolor}

\usepackage{amsmath, amssymb, amsthm,algorithm}
\usepackage[latin1]{inputenc}

\usepackage{latexsym}
\usepackage{graphics,epsfig}
\usepackage{amsfonts}
\usepackage{booktabs}
\usepackage{color}
\usepackage{multirow}
\usepackage[justification=centering]{caption}

\usepackage{graphicx}
\usepackage{adjustbox}
\usepackage{kantlipsum}
\usepackage[caption=false,font=footnotesize]{subfig}
\usepackage{cases}
\usepackage{url}
\usepackage{tabu}
\usepackage{makecell}
\usepackage{soul}

\usepackage{array}
\usepackage{algpseudocode}
\usepackage{mathtools,lipsum,cuted}

\usepackage[justification=centering]{caption}

\usepackage[caption=false,font=footnotesize]{subfig}
\usepackage[T1]{fontenc}
\usepackage{mwe}
\usepackage{subfig}

\newlength{\tempheight}
\newlength{\tempwidth}

\newcommand{\rowname}[1]% #1 = text
{\rotatebox{90}{\makebox[\tempheight][c]{#1}}}

\newcommand{\columnname}[1]% #1 = text
{\makebox[\tempwidth][c]{#1}}

\makeatletter
\def\BState{\State\hskip-\ALG@thistlm}
\makeatother

\newcounter{example}[section]

\usepackage[english]{babel}
\usepackage{verbatim}

\usepackage[english]{babel}

\theoremstyle{plain}

\theoremstyle{remark}

\def\BibTeX{{\rm B\kern-.05em{\sc i\kern-.025em b}\kern-.08em
    T\kern-.1667em\lower.7ex\hbox{E}\kern-.125emX}}
\begin{document}
\title{Towards Optimal Path Allocation for Unreliable
Reconfigurable Intelligent Surfaces
%Unreliability-Aware path Allocation of Virtual Reality over Terahertz Reconfigurable Intelligent Surfaces.\
}

\author{\IEEEauthorblockN{Mounir Bensalem$^{*}$, Anna Engelmann$^{*}$ and Admela Jukan$^{*}$}
\IEEEauthorblockA{$^{*}$Technische Universit\"at Braunschweig, Germany;
\{mounir.bensalem, a.jukan\}@tu-bs.de}a.engelmann.ida@gmail.com

%$^{+}$CARIAD SE, Germany;
%anna.engelmann@cariad.technology

}

\maketitle

\begin{abstract}%\hl{updated} $\surd$\
Terahertz (THz) communications and reconfigurable intelligent surfaces (RISs) have been recently proposed to enable various powerful indoor applications, such as wireless virtual reality (VR). For an efficient servicing of VR users, an efficient THz path allocation solution becomes a necessity. Assuming the RIS component is the most critical one in enabling the service, we investigate the impact of RIS hardware failure on path allocation performance. To this end, we study a THz network that employs THz operated RISs acting as base stations, serving VR users. We propose a Semi-Markov decision Process (SMDP)-based path allocation model to ensure the reliability of THz connection, while maximizing the total long-term expected system reward, considering the system gains, costs of link utilization, and the penalty of RIS failure. The SMDP-based model of the RIS system is formulated by defining the state space, action space, reward model, and transition probability distribution. We propose an optimal iterative algorithm for path allocation that decides the next action at each system state. The results show the average reward and VR service blocking probability under different scenarios and with various VR service arrivals and RIS failure rates, as first step towards feasible VR services over unreliable THz RIS.
\end{abstract}


\section{Introduction}%\hl{updated} $\surd$\\
% Situation, Stat of the art, why people need to address Unreliability-Aware path Allocation of VR over RIS
%
Virtual Reality (VR) applications are envisioned as one of the key technologies that will  advance the human to machine interactions where a user is present and acts in a virtual world \cite{saad2019vision}. In order to satisfy the high demand of data rates that VR applications require, the THz frequency band is currently one of the most promising communications technology to  provide a high quality VR experience. As THz communication is characterized by directional and sensitive to attenuation beams, a new technology called  Reconfigurable Intelligent Surfaces (RIS) is used to enhance the performance of THz wireless communications \cite{bensalem2021benchmarking}.
In an indoor setting, the usage of RIS in THz communications is rather critical, since by controlling the reflective properties of the underlying channels, we can mitigate various THz transmission impairments \cite{wu2021intelligent}. 

\par It is known that RIS elements, also called unit cells or meta-atoms, are vulnerable to failures, which can cause deterioration of the antenna radiation pattern, and in severe cases, the RIS failures can affect the functioning of RIS meta-surface elements \cite{li2021array}. In addition,THz links between users and RISs can be blocked, due to various obstacles. In order to design reliable RIS based networks in practical environments, it is becoming essential to develop path allocation schemes that are aware of RIS meta-surface failures, as well as link failure between RISs and users. The path allocation problem in RIS networks is indeed one of the key problems to solve for any efficient RIS design, where several sub-problems can be investigated, including channel allocation,  power allocation, RIS-to-user assignment, and phase shifts design. Few works studied resource allocation problems, but no work to date considered the problem of path allocation under faulty RIS scenarios. To the best of our knowledge, path allocation of VR over THz RIS, considering the unreliability of RIS devices has not yet been investigated.

%%%%%%%%%%%%%%%%%%%%%%%%%%%%%%%%%%%%%%%%%%%%%%
%%Resolution: how did we solve the problem, what is our novel approach
%%%%%%%%%%%%%%%%%%%%%%%%%%%%%%%%%%%%%%%%%%%%%
%
%%\hl{we can work on intro at the end}
In this paper, we consider path  allocation in the VR applications over RIS  network with controlled access of VR users request by Semi-Markov decision Process (SMDP). We introduce an optimal path allocation scheme to ensure the reliability and maximize the system reward in a set of RISs, whereby a RIS device consists of blocks of meta-surfaces which are vulnerable to failures.  We formulate the problem of RIS-unreliability-aware path allocation as an SMDP model considering multiple RIS devices in an indoor environment used to allocate THz meta-surfaces blocks to VR users. %We considered as service the channel \hl{define service} .
Following the definition of SMDP \cite{puterman2014markov}, the decision is taken at the event occurrence, while an iterative algorithm proposed maximizes the total long-term reward of the VR over RIS system, considering the factors of user's income of the accepted services, the costs for occupying paths, and penalty of RIS and meta-surface failures. We make realistic assumptions on the reference THz network model, where a set of backup RIS devices are used to ensure the reliability of the system. When a RIS or a meta-surface fails, other  blocks of meta-surfaces in the same RIS or in the corresponding backup RIS are allocated to transfer the existing services, which allows the RIS network to operate in a reliable fashion. Numerical results show the average reward and blocking probability with various service arrival and RIS failure rates, and under different network configurations. %Our proposed solution can compute corresponding policies to any arrival rate of service requests to ensure the reliability and to maximize long-term rewards. The proposed scheme % is generally applicable,  dynamic and provides an efficient solution to the path allocation problem. 
We  show that our scheme improves the QoS of THz communication with regard to the arrival rate of service requests and RIS meta-surfaces availability. %FOLLOWING SENTENCE IS MAYBE FOR CONCLUSION... We also show that future work can benefit from a large-scale solution based on reinforcement learning that could be used to solve value iteration algorithm, which is known to grow exponentially.  % FOLLOWING SENTENCE IS FOR RESULT- OR ASSUMPTION-SECTION --> Moreover, when VR users request arrival rate is high, the paths are allocated with the minimum requirements for a VR service to operate, which decreases the blocking probability, and allows more users to join the RIS network overall. %Since each RIS device can be used to provide a connection to various VR users, we show that a shared solution allows for more flexibility to the system. 
The results show that increasing the number of RISs and meta-surfaces does not always improve the performance in terms of long-term reward and service blocking for service arrival rates below a certain threshold. Hence, our model can help dimensioning RIS THz networks in terms of costs and performance based on the expected service rate. 

%
%%%%%%%%%%%%%%%%%%%%%%%%%%%%%%%%
%
%
%
%%%%%%%%%%%%%%%%%%%%%%%%%%%%%%%%%%%%%%%%%%%%%
%
The rest of this paper is organized as follows. Section \ref{sec:0} discusses the related work. Section \ref{sec:1} presents the SMDP based path allocation model in indoor THz network, with unreliable RIS. Section  \ref{sec:SMDP} describes the proposed optimal iterative-based solution. Section \ref{sec:results} evaluates the performance. We conclude the paper in Section \ref{sec:conclusion}.
\section{Related Work}\label{sec:0}
%
%% works on VR and RIS
VR communication has been recently investigated in several seminal works \cite{chen2018virtual, %kasgari2019human,
 chaccour2020risk }, focusing on studying the VR Quality of Service (QoS). 
In \cite{chen2018virtual}, a VR model was studied that detect the tracking and delay components of VR QoS. %The work in \cite{kasgari2019human} analysed  the spectrum resource allocation problem, considering a brain-aware QoS constraint.
 In \cite{chaccour2020risk} the downlink of VR requests over RISs network, operating over the terahertz (THz) frequency bands, was considered, and a solution was proposed to the problem of associating RISs to virtual reality users.

%% Importance of failure and related references 
Due to the importance of hardware failures, the baseband complex received signal from a RIS element has been modeled in \cite{wang2021doppler}, considering  faulty reflecting elements of RIS. Thus, in order to design reliable RIS based networks in practical environments, it is essential to develop resource allocation schemes aware of RIS array failures. The optimal number of reflecting elements in a RIS has been studied theoretically in \cite{zappone2020optimal} to maximize the transmission rate in a point-to-point link. %, emphasizing the importance of RIS configuration (e.g., path allocation scheme), and the resulting overhead due to channel estimation.
  Motivated by this work, which concluded that new, sophisticated resource allocation schemes are needed, we use a model of a VR over THz RIS network under the similar failure assumptions to solving path allocation problem. 
%
%resource allocation problem and related work
Recent works studied the resource allocation problem in THz RIS networks specifically. Paper \cite{huang2020holographic} studied different types of active  RISs (active and passive), considering their hardware architectures, operation modes, and applications in communications, and also highlighted the need of developing efficient resource allocation solutions, which is still an open challenge. % In \cite{di2020hybrid} a downlink multi-user multi-input single-output (MISO) system was studied, where a RIS is used to support a base station to ensuring a reliable communication when the signal is affected by obstacles. The resource allocation problem is tackled by an iterative algorithm based on alternating optimization, in order to solve the sum-rate maximization problem,  addressed by optimizing the digital beam-forming at the base station and the discrete phase shifts at the RIS. The main focus of the work was the rate optimization over a signal RIS component, under the assumption that RIS devices are reliable, and a user is assigned to a single channel through a single path. 
In \cite{chaccour2020risk}, a virtual reality network was considered and a solution was proposed to the problem of associating RISs to virtual reality users operating over the terahertz frequency bands. In particular, the paper formulates a risk-based framework based on the entropic value-at-risk and optimize the transmission rate and reliability. A single RIS device was considered and the formulated problem aims at achieving higher order statistics of the queue length, in order to guaranteeing continuous reliability.  % Lyapunov optimization, deep reinforcement learning, and recurrent neural network were adopted to solve the optimization problem.
 We adopt a similar VR over RIS model as in \cite{chaccour2020risk}, however our focus is not on optimizing the transmission rate by adjusting the queue length. Instead, we assume the existence of multiple RISs that can provide the same set of VR users,  as well as the possibility to allocate multiple paths for the same VR user request, while considering the failure possibility of RISs and meta-surfaces. 
%
%
%
%A few other related works are notable including \cite{huang2019reconfigurable}, that focused on energy-efficient design for transmit power allocation and the phase shifts of the surface reflecting elements.   In \cite{jung2021optimality}, an optimal scheme was proposed that maximizes the system sum-rate using a modulation scheme compatible with RIS system and a resource allocation scheme that control the transmission power and  modulation. The studied system has not considered  path allocation, considering the user arrival rates, service rates, or the faulty RISs. 
%
%Paper \cite{zuo2020resource} studied the resource allocation problem in RIS systems focusing on power allocation and proposed a resource allocation framework in multi-cell downlink RIS-non-orthogonal multiple access (NOMA) networks, where a joint optimization problem was solved to maximize the achievable sum-rate of user association, sub-channel assignment, power allocation, phase shifts design, and decoding order determination; a single RIS was considered assuming that RIS devices cannot fail. In \cite{li2019joint}, joint active and passive beamforming problem and the user-RIS association problem was investigated in a  multi-RIS assisted multi-user communications. The authors  mapped the problem of max-min signal-to-interference-plus-noise ratio (SINR) into a user-RIS association problem, and proposed greedy search algorithm as a solution, tuning  the RIS to serve a certain user.
%
%% channel allocation in other context with SMDP
\par It should be noted that path allocation problem has been well investigated in other contexts such as in vehicular networks. In  \cite{li2017smdp}, an SMDP-based path allocation scheme was proposed in vehicular ad-hoc networks to maximize the overall system rewards, while servicing user requests through roadside units. The paper concluded that due to the high mobility of users, an SMDP policy for the path allocation problem can improve the QoS. Motivated by this work, we could see the VR over RIS network as a cognitive-enabled vehicular ad hoc networks, where VR users can be mapped to vehicle users, RISs could replace the roadside units (RSUs), and the allocation of meta-surface blocks can be seen as path allocation, which is our approach in this paper. SMDP-model was also proposed in \cite{li2020spectrum} to solve path allocation problem in order to maximizing throughput in the context of vehicular networks. 
%
%%Advantage compared to other methods
The related work is primarily based on linear programming which is known to be an NP hard problem, and that it works for off-line processing, whereas our proposed model works for online channel assignment to users using an iterative solution, that assumes the knowledge about some system statistics, such as the RIS failure, service arrival, and service departure.  % limitations 
Moreover, our approach assumes that the system statistics are known, and that the RIS configuration such as phase shift and power control are managed by other algorithms independently. Thus, our approach is focused on path allocation only, as a function of VR user's THz channel request behavior. We also do not consider the dependency between RIS failures, which is a limitation, as is the intrinsic limitation of SMDP to scale to larger systems.

\section{SMDP based Channel Allocation of VR over THz RIS}\label{sec:1} %\hl{updated}$\surd$
%%  Add reliability issues  %%
%\hl{Hypothesis:}\\
%\begin{itemize}
%    \item each RIS has several meta-surfaces
%    \item a user can be served by multiple RISs
%    \item a RIS device can fail with low probability
%    \item RIS controller allocate and configure  meta-surfaces for users
%\end{itemize}
%%We present a model where IoT end-devices are sending service request on various classes of services with random arrival and departure rate.

In this section, we model a RIS-based wireless reference THz network in
an indoor area, provide the channel model, analyze the channel states and actions,  transmission probabilities, as well as rewards analysis.

\begin{figure*}
 \centering\vspace{-0.5 cm}
   \includegraphics[scale=0.5]{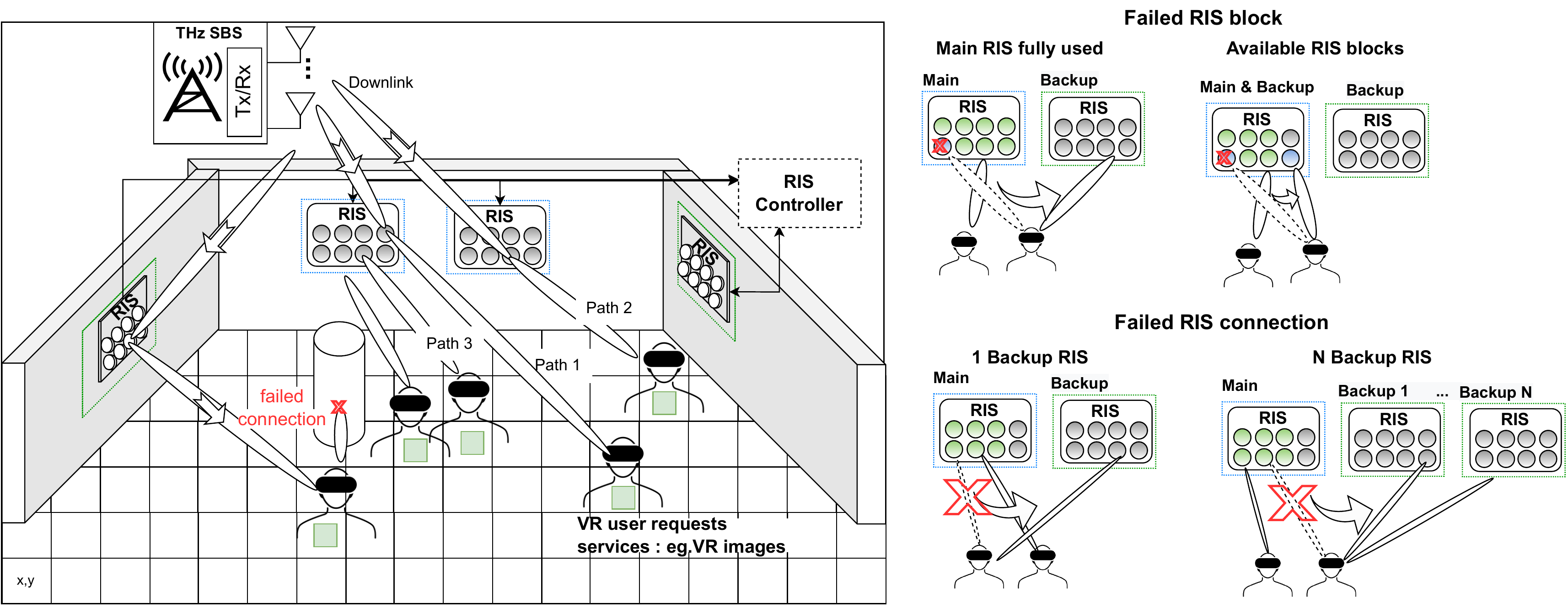}%arch_thz_new1.pdf}
 \caption{The reference indoor THz RIS network.}
\label{fig:arch}
\vspace{-0.5 cm}
\end{figure*}
\subsection{The reference network and assumptions} 
%%%%%%%%%%%%%%%%%%%%%%%%%%%%%%%%%%%%%%%%
%%%%%%%%%%%%%%%%%%%%%%%%%%%%%%%%%%%%%%%%
The reference THz network is illustrated in Fig.  \ref{fig:arch}. The THz network consists of a set of RISs, small base station (SBS) operating over THz frequency, RIS controller, and VR users distributed in an indoor environment. RISs are placed in an indoor environment where they can be reached by users in certain areas based on the distance, direction and the existence of obstacles. The users are assumed to be mobile and can have different locations in the indoor environment. The VR user's end-devices can send several service requests through the RIS network. Each mobile wireless VR user is serviced by a main RIS, using a certain allocated blocks of meta-surfaces. We assume that the block of  meta-surfaces can be non-functional due to hardware problems, e.g., due to dirty surfaces or failure. Thus, the RIS controller allows to re-allocate service request to another available meta-surfaces of the same RIS, or meta-surfaces of a backup RIS.
 The SBS is responsible for receiving requests from VR users and service them via RIS components. The uplink of VR service requests is adopting an ultra reliable low latency communications (URLLC) scheme, similar to \cite{chaccour2020risk}. In this study, we consider the downlink of the RIS-based network only, where each RIS is covering a part of the indoor environment and can provide THz communication channel to VR users. The end-devices of VR users can send several service requests through the RIS network. \\
Each RIS can reflect the beams received from a main THz SBS, which generally uses a MIMO system. % The information sent from the SBS  is encoded in the phase of the signal reflected by the reconfigurable elements that compose the RIS. 
A RIS contains a very large number of passive reflective elements called meta-surfaces, capable of reconfiguring properties of electromagnetic (EM) waves impinging upon them. We assume that all RISs are connected to the \textit{RIS Controller} responsible for channel tuning by changing the phase of the reflected signal \cite{huang2019reconfigurable}. We assume that a RIS contains several blocks of meta-surfaces, configured to create a path from the SBS to a user. These blocks can be allocated based on the system availability. In order to ensure the reliability of the system, we pre-reserve a backup RIS to each operating RIS. We assume that services can be transferred from the operating RIS to the backup RIS when  operating RIS or a single block of meta-surfaces or a THz connection to a RIS fails.  Figure \ref{fig:arch} illustrates three failure scenarios. First, when a block of meta-surfaces fails, we assume the failure of the block will affect at most one user, while all other RIS blocks remain unaffected. To insure the reliability, we first check whether the main operating RIS has available meta-surfaces for re-allocation, otherwise connection from the main RIS is transferred to an available backup RIS. When an entire RIS fails, all connections will be affected and transferred to a backup RIS. The same recovery strategy is applied, is a THz connection between RIS and VR users fails, as shown in Fig.  \ref{fig:arch}. %\hl{It is also important to know, if the backup channel is pre-reserved or can be selected on the fly> What is the difference ? which one is better and fit the model ?}

Without loss of generality, we assume that the RIS network can include heterogeneous RIS devices, of various sizes and containing different number of blocks of meta-surfaces, and thus servicing different number of user through multiple paths. 
 When a VR service requested by VR user is detected, the RIS controller accepts or rejects it based on the availability of paths, i.e the RIS resources.
In this paper, we assume that the THz link between the VR user and the RIS can be blocked and fail with a certain probability, and the link between RIS and the THz SBS is reliable and cannot be disrupted by obstacles. A VR user-RIS failure caused by an obstacle will result in a failure of the  THz connection and  creates a delay, which affects the QoS.  We assume that a RIS and a meta-surface can face a hardware failure, and propose as next a novel path allocation scheme.
%\hl{PARAGRAPH IS NOT CLEAR, NEEDS TO BE IMPROVED In this paper, we assume that the THz link between the VR user and the RIS can be blocked and fail with a certain probability, and the link between RIS and the THz SBS is reliable and cannot be disrupted by obstacles. A VR user-RIS failure caused by an obstacle  will result in a failure of the transmission channel, thus in the allocated path,  and  creates a delay, which affects the QoS.  We assume that a RIS and a meta-surface can face a hardware failure, and for the downlink connection we propose a novel path allocation scheme.}

%\hl{talk about reliability}\\
\subsection{Channel Model}
This model focuses on the THz link between VR users and the RIS, where the VR user is mobile. The user movements can cause signal blocking due to self blocking, or due to other mobile or fixed obstacle in the indoor environment, as described in \cite{chaccour2020risk}. We assume the time slots indexed by $t$ have a period duration $\tau_0$. We denote by $L_{r,n,u,t}$  a binary random variable, which is equal to $1$ when the Line-of-Sight (LoS) link between the meta-surface $n$ of the RIS $r$ and the user $u$ at time slot $t$ is available and 0 otherwise. The corresponding random channel gain is defined as in \cite{chaccour2020risk} as:

\begin{equation}
 h_{r,n,u,t}\!\! =\!\! \begin{cases}
     \left( \frac{c}{4\pi f d_{r,u,t}} \right)^2  e^{-k(f)d_{r,u,t}}  ,& \text{if } P(L_{r,n,u,t}=1) ,\\ 
     0,& \text{if } P(L_{r,n,u,t}=0) .
    \end{cases} 
\end{equation}
where $c$ is the speed of light in the vacuum, $d_{r,u,t}$ is the distance between RIS $r$ and the VR user at time slot $t$, $f$ is the operating frequency, $k(f)$ is the overall molecular absorption coefficients of the medium at THz band, available from HITRAN database \cite{rothman2009hitran}. \\
The transmission rate between RIS $r$ and the VR user $u$ at time slot $t$ is given by:
\begin{equation}
\begin{split}
c_{r,u,t} & = W\cdot \log_2 \\
&\left(  1+  \frac{P_{RIS}  h_{r,n,u,t}\sum_{n=1}^{N} |e^{j(\phi_{r,n,u,t} - \psi_{r,n,u,t})} |^{2} L_{r,n,u,t}}{\mathcal{N}(d_{r,u,t},p,f)} \right)\\
\end{split}
\end{equation}
where $N$ is the number of meta-surface blocks in a RIS, $W$ is the total bandwidth, $P_{RIS}$ is the transmit power related to each RIS, $\psi_{r,n,u,t}$ is the phase shift of meta-surface element $n$ of RIS $r$ with respect to user $u$ at time slot $t$, $\phi_{r,n,u,t}$ is the phase shift of the channel between meta-surface element $n$ of RIS $r$ with respect to user $u$ at time slot $t$, $\mathcal{N}(d_{r,u,t},p,f)= N_0 + \sum_{r=1}^{R}P_{RIS} A_0d_{r,u,t}^{-2}(1-e^{-K(f)d_{r,u,t}}) $,   $N_0=\frac{W\lambda^2}{4\pi}k_B T_0$, $T_0$ is the temperature (Kelvin), $k_B$ is the Bolzmann constant, $A_0=\frac{c^2}{16\pi^2 f^2}$ \cite{jornet2011channel}.\\

The obtained transmission rate can be used to calculate the service rate of a specific VR service $s$, given by:
\begin{equation}\label{service_rate}
C_{r,u,t,s}=\frac{c_{r,u,t}\tau_0}{O_s}
\end{equation}
where $\tau_0$ is the duration between two discrete time slots,  $O_s$ is the size of the object related to the service $s$, such as the size of a VR image.

\subsection{Problem Formulation}

We consider a RIS-based THz network including a set of RISs, i.e., a set of operating RISs $\mathbf{R}= \{ r_1,...,r_i,...r_{R} \}$ and each RIS with index $i$ is assigned to a different RIS called a backup RIS, the set of backup RISs is $\mathbf{B}= \{ b_1,...,b_i,...b_{R} \}$, managed by the RIS controller. We study the channel allocation problem regarding different RIS components, where VR users are requesting service connections. 
The RIS $r_i$ and $b_i$ contain $N(i)$ meta-surface blocks that can be allocated to create THz paths.  A service request can occupy $k\in \{1,...,K\}$ block based on the resource availability, where $K$ represents the maximum number of blocks a RIS can allocate to a single service. %, whereby $K\leq min_{\substack{\forall i: i\in[1,R]}}\{N(i)\}$. 
%For any service, all RIS  has enough maximum capacity $N(i)$ to satisfy its capacity requirements. 
We model the arrival and service processes of VR service requests from a user as a Poisson process with mean rates $\lambda_{s}$, and $\mu_s$, respectively, where $\mu_s$ can be concluded from eq. (\ref{service_rate}) for different services. \\
\subsection{System States}
%%%%%%%%%%%%%%%%%%%%%%%%%%%%%%%%%%%%%%%%%%%%%

%%%%%%%%%%%%%%%%%%%%%%%%%%%%%%%%%%%%%%%%%%%%%

The system state $s$ represents the number of VR service requests with different number of allocated block $k$ in  a RIS  $r_{i}$,  the availability  of the RIS and  the next  event that can happen in the system:
\begin{equation}
\begin{split}
S=\{s|s=& (\Delta,  X, e)\},
\end{split}
\end{equation}

where a set $\Delta=\delta^{r_1},...,\delta^{r_i},...,\delta^{r_R}$ collects all subsets $\delta^{r_i}$, which describe a state $s$ and indicate a number of VR services requested 
$\delta^{r_i}=\{\delta_{1}(r_i),...,\delta_{k}(r_i),...,\delta_{K}(r_i) \}$. The variable $\delta_{k}(r_i)$  denotes the number of  VR services allocated with $k$ blocks in the main RIS  $r_i$.
A set $X=\{X^{r_1},..., X^{r_R} \}$ describes the availability of RIS devices and indicates the number of available meta-surface blocks: $X^{r_i}= N(i)$ if all the elements of the RIS $r_i$ are available, $X^{r_i}= N(i) - j$ if $j$ elements are failed  and $X^{r_i}= 0$ if the all elements are failed or the RIS hardware fails. 
The variable $e$ describes an event that occurs in the VR over THz RIS network, such as $e = \{Ar, D^r,  F^{m}, Re^{m}\}$, where a set $Ar$ 
denotes  the arrival  of any  VR service request, a set $D^r=\{D^{r_1},...,D^{r_i},...,D^{r_R}\}$, and a subset $D^{r_i}=\{D^{r_i}_{1},...,D^{r_i}_{k},...,D^{r_i}_{K}\}$ collects the set of departure events of a VR service from a main RIS  $r_i$, and $D^{r_i}_{k}$ defines the departure  of a VR service allocated in $k$ blocks from a RIS  $r_i$, a set $F^{m}=\{F_{1}^{m},...,F_{i}^{m},...,F_{R}^{m}\}$ and a variable  $F_{i}^{m}$ define the failure process of  a meta-surface block in the RIS  $r_i$,   a set $Re^{m}=\{Re_{1}^{m},,...,Re_{i}^{m},...,Re_{R}^{m}\}$ and a variable $Re_{i}^{m}$ define the return of a path/meta-surface block in the RIS $r_i$ to working state after a failure.

%The path allocation scheme in RIS network has the following capacity constraints:
%\begin{equation}\label{CapConst}
%\forall i\in [1,R]:   \sum_{k=1}^{K} k\cdot \delta_{k}(r_i) \leq N(i), 
%\end{equation}

\subsection{Actions}
The system controller has several possibilities of actions a to take when it receives a service request, whether to accept or reject it. The action space $A(s)$ is described as follows:
\begin{equation}
A(s)= \begin{cases}
      \{0, (i,k)\} ,\;& e\in Ar, \\ & k\in \{0, 1, 2,..., K\},i \in \{ 1,...,R\} \\
     -1, & e\in \{D^r, Re^{m}_i\}\\
     (-2, T), & e\in  F^{m}_i
    \end{cases}
\end{equation}
% in next paper extension or improvement, we can propose migrating services from failed channels to another channel, and consider this here in the action 
where $a(s)=(i,k),\forall k\in \{1,...,K\}\forall i\in \{1,...,R\}$ when a VR service request is accepted and    $k$ blocks are allocated in RIS $r_i$, $a(s)=0$ denotes the action of rejecting a service request. When a service completes and departs the RIS system, a RIS  node or meta-surface  block returns into the system after a failure, no action is required, and the controller needs only to update the system state, we represent the action as $a(s)=-1$. When a RIS  fails, all the allocated blocks  are transferred to the backup RIS, and when a meta-surface block fails, a corresponding block will be allocated in the same RIS if resources are available or in a backup RIS otherwise, we denote the information update and the transfer action as $a(s)=(-2,T)$, where T is the transfer vector from current RIS state to re-allocation state. %should add actions for this in the future
%\begin{defn}
%we define a mapping $\psi(.)$ in $\mathbb{N}$ as: $\psi(x)=0$ if $x=0$, and $\psi(x)=1$ if $x>0$.
%\end{defn}

\subsection{Transition Probabilities}

%%%%%%%%%%%%%%%%%%%%%%%%%%%%%%%%%
We assume that the time period between two continuous decision epoches follows an exponential distribution and denoted as $\tau (s,a)$, given the current state
$s$ and action $a$. Thus the mean rate of events for a specific state $s$ and action $a$ denoted as $\gamma(s,a)$, is the sum of the rates of all events in the RIS system, which is expressed as follows:
\begin{strip}
\begin{equation}
\begin{split}
& \forall i \in \{1,...,R\}, \forall k \in \{1,...,K\}: \text{    } \tau(s,a)=  \gamma(s,a)^{-1}= \\  &
 \begin{cases}
       \lambda_s +  \Lambda^{m} + \Theta^{m} + \Theta  + k \mu_s  ,\;& e= Ar, a=(i,k), \\
     \lambda_s + \ \Lambda^{m} + \Theta^{m}   + \Theta , & e= Ar, a=0, \\
     \lambda_s +  (\Lambda^{m}-\lambda^{m})  +  (\Theta^{m} +\mu^{m})  + \Theta ,  & e=Re_{i}^{m}, a=-1, \\
     \lambda_s +  \Lambda^{m} +  \Theta^{m}  + \Theta - k\mu_s,  & e=D_k^{r_i}, a=-1, \\
      \lambda_s +  (\Lambda^{m} +\lambda^{m}) + (\Theta^{m}-\mu^{m})+ \Theta   , & e=F_{i}^{m}, a=(-2,T),
    \end{cases}
\end{split}\label{eq:tau}
\end{equation}
\end{strip}

where $\lambda_s$ is the  arrival rate of  VR service requests. % $\lambda^{r}$ and $\mu^{r}$ denotes the arrival rate  and the departure rate (failure) of a RIS block,  $\Lambda^{r}=\lambda^{r} \sum_{i=1}^{R} (1-X^{r_i})$ is the arrival rate of non available RISs, $\Theta^{r}=\mu^{r} \sum_{i=1}^{R} X^{r_i}$ is the failure rate of available RIS blocks,
  $\Lambda^{m} =\lambda^{m}\sum_{i=1}^{R} (N(i)-X^{r_i})$ is the arrival rate of non available RIS blocks, and $\Theta^{m}=\mu^{m}  \sum_{i=1}^{R} N(i)$ is the failure rate of available RIS blocks.   

When an arriving service request is rejected, or a RIS  or a RIS block returns to system, the total number of blocks  allocated in RISs is $\sum_{i=1}^{R} \sum_{k=1}^{K}\delta_{k}(r_i)$, so the departure rate of a VR service in RIS system  is $\Theta=\sum_{i=1}^{R} \sum_{k=1}^{K}k\delta_{k}(r_i) \mu_s$. When a service request is accepted and $k$ blocks are allocated, one more service is added to the system, thus the departure rate becomes $\Theta + k \mu_s$.  When a departure of a VR service  from a RIS $r_i$ allocated in  $k$ blocks occurs, the departure rate becomes  $\Theta - k\mu_s$. When a RIS block fails, the arrival rate and failure rate of the RIS block in the system are adjusted such that the failed block can return in the future and cannot fail again while it is already failed. All services admitted in the case of a  failed RIS block are transferred to a backup RIS block, so it should  be accounted in the departure rate, which ensures the system reliability, and thus the number of existing services remains equal to $\Theta$. %When a RIS fails,   the arrival rate and failure rate of RIS in the system are also adjusted, and all services are transferred to the backup RIS, thus the service rate remains the same.
%%%%%%%%%%% Transition Probability %%%%%%%%%%
The transition probability in our markov decision model from state $s$ to state $s'$ when an action $a$ is selected is denotes as $p(s'|s,a)$, which can be determined under different events.
%\hl{Mounir, you also need to describe what each state mean! At least one sentence per state!}
\begin{itemize}
\item State $s=(\Delta,  X, Ar)$, and $a=0$. %This state describes the system in terms of number of channels allocated in a RIS $r_i$, RIS and RIS meta-surfaces availability, and the next event, which is in this case a service arrival. The service arrival event can have two actions accept or reject. The following equation shows the transition probability when the service is blocked, where the number of channels allocated in the RISs and in the backup RISs, and RIS and RIS meta-surfaces  availability remain the same, and possible events can occur in the future.  
\end{itemize}

\begin{equation}
\begin{split}
&p(s'|s,a)= %\\= &
 \begin{cases}
     \frac{\lambda_{s}}{\tau(s,a)}   &   s'=(\Delta,  X, Ar) \\
      \frac{k\delta_{k}(r_{i}) \mu_s}{\tau(s,a)}   &  s'=(\Delta,  X, D_{k}^{r_{i}}) \\
        \frac{\lambda^{m}}{\tau(s,a)}   &   s'=(\Delta,  X, Re_{i}^{m}) \\
         \frac{\mu^{m}}{\tau(s,a)}   &   s'=(\Delta,  X, F_{i}^{m}) \\
    \end{cases}\label{eq:tr1}
\end{split}
\end{equation}

\begin{itemize}
\item State $s=(\Delta, X, Ar) $, $a=(i,k)$, and $k\geq 1$. %Similar to the previous case, this state describes the system in terms of number of channels allocated in all RISs, RIS and RIS meta-surfaces availability, and an event of service arrival, when the service is accepted and $k$ channels are allocated in RIS $r_i$.
\end{itemize}
\begin{equation}
\begin{split}
&p(s'|s,a)= %\\= &
 \begin{cases}
     \frac{\lambda_{s}}{\beta(s)\tau(s,a)}   &   s'=(\widehat{\Delta},  X, Ar) \\
      \frac{k(\delta_{k}(r_{i})+1) \mu_s}{\beta(s)\tau(s,a)}   &  s'=(\widehat{\Delta},   X, D_{k}^{r_{i}}) \\
      \frac{k'\delta_{k'}(r_{i'}) \mu_s}{\beta(s)\tau(s,a)}   &  s'=(\widehat{\Delta},   X, D_{k'}^{r_{i'}}) \\ & k'\neq k, i'=i || k'=k, i'\neq i\\
       \frac{\lambda^{m}}{\beta(s)\tau(s,a)}   &   s'=(\widehat{\Delta},   X, Re_{i'}^{m}) \\
       \frac{\mu^{m}}{\beta(s)\tau(s,a)}   &   s'=(\widehat{\Delta},   X, F_{i'}^{m}) \\
    \end{cases}\label{eq:tr2}
\end{split}
\end{equation}

where $\beta(s)=\sum_{i=1}^{R} X^{r_i}$ denotes the number of RIS blocks that are available at the state $s$, and $\widehat{\Delta}=\delta^{r_1},...,\delta^{r_i}+ I_k,...,\delta^{r_R}$, $I_k$ denotes a vector having $K$ elements, with k-th element equal to 1 and 0 for the others.

\begin{itemize}
\item State $s=(\Delta,   X,  D_{k}^{r_{i}})$, $a=-1$. %This state describes the system in terms of number of channels allocated in all RISs, RIS and RIS meta-surfaces availability, and an event of  service departure with $k$ allocated channels from the main RIS $r_i$. 
\end{itemize}

\begin{equation}
\begin{split}
&p(s'|s,a)= %\\= &
 \begin{cases}
     \frac{\lambda_{s}}{\tau(s,a)}   &   s'=(\widehat{\Delta},   X, Ar) \\
      \frac{k(\delta_{k}(r_{i})-1) \mu_s}{\tau(s,a)}   &  s'=(\widehat{\Delta},   X, D_{k}^{r_{i}}) \\
      \frac{k'\delta_{k'}(r_{i'}) \mu_s}{\tau(s,a)}   &  s'=(\widehat{\Delta},   X, D_{k'}^{r_{i'}}) \\ & k'\neq k, i'=i || k'=k, i'\neq i\\
        \frac{\lambda^{m}}{\tau(s,a)}   &   s'=(\widehat{\Delta},   X, Re_{i'}^{m}) \\
        \frac{\mu^{m}}{\tau(s,a)}   &   s'=(\widehat{\Delta},   X, F_{i'}^{m}) \\
    \end{cases}\label{eq:tr3}
\end{split}
\end{equation}
where $\widehat{\Delta}=\delta^{r_1},...,\delta^{r_i}- I_k,...,\delta^{r_R}$.
%
%\begin{itemize}
%\item State $s=(\Delta,   X, Re^{r_{i}})$, $a=-1$. %This state describes the system in terms of number of channels allocated in all RISs, RIS and RIS meta-surfaces availability, and the next event, which is  a return of a RIS  $r_i$ to working state after a failure, where $X^{r_i}=0$.
%\end{itemize}
%\begin{equation}
%\begin{split}
%&p(s'|s,a)= %\\= &
% \begin{cases}
%     \frac{\lambda_{s}}{\tau(s,a)}   &   s'=(\Delta,   \widehat{X}, Ar) \\
%      \frac{k\delta_{k}(r_{i'}) \mu_s}{\tau(s,a)}   &  s'=(\Delta,   \widehat{X}, D_{k}^{r_{i'}}) \\
%    &  i'\neq i \\
%        \frac{\lambda^{m}}{\tau(s,a)}   &   s'=(\Delta,   \widehat{X}, Re_{i}^{m}) \\
%        \frac{\mu^{m}}{\tau(s,a)}   &   s'=(\Delta,   \widehat{X}, F_{i}^{m}) \\
%    \end{cases}\label{eq:tr5}
%\end{split}
%\end{equation}
%where  $\widehat{X}^{r_i}= N(i) $
\begin{itemize}
\item State $s=(\Delta,   X, Re_i^{m})$, $a=-1$, where $X^{r_i}=N(i)-j$, and $j$ is the number of failed RIS blocks. %This state describes the system in terms of number of channels allocated in all RISs, RIS and RIS meta-surfaces availability, and the next event, which is  a return of a meta-surface in RIS $r_i$ to the working state after a failure, where $X^{r_i}=N(i)-j$, and $j$ is the number of failed meta-surfaces.
\end{itemize}
\begin{equation}
\begin{split}
&p(s'|s,a)= %\\= &
 \begin{cases}
     \frac{\lambda_{s}}{\tau(s,a)}   &   s'=(\Delta,   \widehat{X}, Ar) \\
      \frac{k\delta_{k}(r_{i'}) \mu_s}{\tau(s,a)}   &  s'=(\Delta^{R},   \widehat{X}, D_{k}^{r_{i'}}) \\
    &  i'\neq i \\
        \frac{\lambda^{m}}{\tau(s,a)}   &   s'=(\Delta,   \widehat{X}, Re_{i}^{m}) \\
         \frac{\mu^{m}}{\tau(s,a)}   &   s'=(\Delta,   \widehat{X}, F_{i}^{m}) \\
    \end{cases}\label{eq:tr6}
\end{split}
\end{equation}
where  $\widehat{X}^{r_i}= N(i)-j+1 $
%\begin{itemize}
%\item State $s=(\Delta,   X,  F^{r_i})$, $a=(-2, T)$. % This state describes the system in terms of number of channels allocated in all RISs, RIS and RIS meta-surfaces availability, and the next event, which is the failure process of  a RIS  $r_i$, where $X^{r_i}=N(i)-j \geq 1$. The vector $T$ transfers the allocated channels in $r_i$ to $b_i$.
%\end{itemize}
%\begin{equation}
%\begin{split}
%&p(s'|s,a)= %\\= &
% \begin{cases}
%     \frac{\lambda_{s}}{\tau(s,a)}   &   s'=(\widehat{\Delta},   \widehat{X}, Ar) \\
%      \frac{k\delta_{k}(r_{i'}) \mu_s}{\tau(s,a)}   &  s'=(\widehat{\Delta},   \widehat{X}, D_{k}^{r_{i'}}) \\
%    &  i'\neq i \\
%       \frac{\lambda^{r}}{\tau(s,a)}   &   s'=(\widehat{\Delta},   \widehat{X}, Re^{r_i}) \\
%        \frac{\lambda^{m}}{\tau(s,a)}   &   s'=(\widehat{\Delta},   \widehat{X}, Re_{i}^{m}) \\
%        \frac{\mu^{r}}{\tau(s,a)}   &  s'=(\widehat{\Delta},   \widehat{X}, F^{r_{i'}}) \\
%    &  i'\neq i \\
%         \frac{\mu^{m}}{\tau(s,a)}   &   s'=(\widehat{\Delta},   \widehat{X}, F_{i'}^{m}) \\
%    &  i'\neq i \\
%    \end{cases}\label{eq:tr7}
%\end{split}
%\end{equation}
%where  $\widehat{X}^{r_i}= 0$, $\widehat{\Delta}=\Delta - T$, $\widehat{\Delta^{B}}=\Delta^{B} + T$, $T=0,...,0,\delta^{r_i},0,...,0$, i.e. $\widehat{\Delta}=\delta^{r_1},...,\delta^{r_{i-1}},0,\delta^{r_{i+1}},...,\delta^{r_R}$.
\begin{itemize}
\item State $s=(\Delta,   X,   F_{i}^{m})$, $a=(-2, T)$,  where $X^{r_i}=N(i)-j \geq 1$,  The vector $T$ transfers one or more  allocated RIS blocks from $r_i$ to the available RIS. % This state describes the system in terms of  number of channels allocated in all RISs, RIS and RIS meta-surfaces availability, and the next event, which is the failure process of  a RIS  meta-surface in the RIS $r_i$, where $X^{r_i}=N(i)-j \geq 1$. The vector $T$ transfers one or more  allocated channels from $r_i$ to the backup RIS $b_i$.
\end{itemize}
\begin{equation}
\begin{split}
&p(s'|s,a)= %\\= &
 \begin{cases}
     \frac{\lambda_{s}}{\tau(s,a)}   &   s'=(\widehat{\Delta},   \widehat{X}, Ar) \\
      \frac{k\delta_{k}(r_{i'}) \mu_s}{\tau(s,a)}   &  s'=(\widehat{\Delta},   \widehat{X}, D_{k}^{r_{i'}}) \\
    & \text{if}\;\; X^{r_i}\leq k \;\;\text{then:}\; i'\neq i \\
        \frac{\lambda^{m}}{\tau(s,a)}   &   s'=(\widehat{\Delta},   \widehat{X}, Re_{i}^{m}) \\
    &  i'\neq i \\
         \frac{\mu^{m}}{\tau(s,a)}   &   s'=(\widehat{\Delta},   \widehat{X}, F_{i'}^{m}) \\
    & \text{if}\;\; X^{r_i}= 1\;\;\text{then:}\; i'\neq i \\
    \end{cases}\label{eq:tr8}
\end{split}
\end{equation}
where  $\widehat{X}^{r_i} = N(i)-j -1$, $\widehat{\Delta}=\Delta - T$,  $T$ is determined using Algorithm. \ref{alg:FSTValgorithm}. 
\begin{algorithm}
\caption{Finding Service Transfer Vector }
\label{alg:FSTValgorithm}
\begin{algorithmic}[1]
\State \textbf{Input: } $s=(\Delta,   X,   F_{i}^{m})$
\State \textbf{Initialization:} $T=T^{1},...,T^{R}$, $T^{i}=\{T^{i}_1,..,T^{i}_k,..,T^{i}_K\}$,\\$T^{i}_k=0, \; \forall i\in \{1,...,R\}, \forall k\in \{1,...,K\}$.
\If {$\widehat{X}^{r_i} = 0$}  $k=1$
\While{ $\delta_{k}(r_i)= 0$ } 
%\State $T^{i}_k = 0$
\State $k=k+1$
\EndWhile
\State $T^{i}_k = -1$,  $T^{b_i}_k=1$
\EndIf
\State \textbf{Return} $T$
\end{algorithmic}
\end{algorithm}
\subsection{Rewards}
Given the system state $s$ and the corresponding action $a$,  the system reward of the VR over THz RIS system is denoted by
\begin{equation}
r(s, a)=w(s,a) - g(s,a)
\end{equation}
where $w(s,a)$ is the net lump sum incomes of VR users at the  state $s$ when action $a$ is taken and an event $e$ occurs, and $g(s,a)$ is the expected system costs.
\begin{equation}
\begin{split}
w(s,a)= 
 \begin{cases}
       R_k  & e= Ar, a=(i,k)\\
        0 & e= Ar_k, a=0\\
         0  & e= Re_{i}^{m}, a=-1\\
       0  & e=D_{k}^{r_{i}}  a=-1 \\
       -\varepsilon  \sum_{k=1}^{K} k T_{k}^{i}  & , e = F_{i}^{m}  a=(-2,T)  
    \end{cases}
\end{split}
\end{equation}
where the variable $R_k=Q-\frac{Z}{k}$ denotes the reward of
the RIS system for  accepting of the requested service and allocating $k$ blocks. $Q$ denotes the income reward from VR user satisfaction, $\frac{Z}{k}$ denotes the transmission cost of occupying $k$ blocks.  The constant $\varepsilon$ denotes the penalty of a meta-surface block failure.  When a meta-surface block fails, a number of services are transferred to the backup RIS, and thus we consider a penalty proportional to the number of re-allocated blocks as  $-\varepsilon  \sum_{k=1}^{K} k T_{k}^{i}$.  In our work, we don't penalize rejected services, neither reward accomplished  services or returned RISs or returned blocks. 

The expected system cost $g(s,a)$ is defined as: 
\begin{equation}
g(s,a)= c(s, a)\cdot \tau (s, a)
\end{equation}
where $\tau(s,a)$ is the expected service time defined by eq. (\ref{eq:tau}) from the  state $s$ to the next state in case that action $a$ is chosen and $c(s,a)$ is the service holding cost rate when the RIS system is in state $s$ in case that action $a$ is selected. Furthermore, $c(s,a)$ can be described by the number of occupied blocks in the RIS system, as follows: 
\begin{equation}
c(s,a)=\sum_{i=1}^{R}\sum_{k=1}^{K}c\cdot k\cdot \delta_{k}^{r_i}
\end{equation}
where $c$ represents the utilization cost of a block unit.
%\begin{figure*}
% \centering
%   \includegraphics[scale=0.45]{figures/transitiondiagram1node1service.pdf}
% \caption{State transition diagram of an edge computing system with one edge node and one service class (Pr=1, N(Pr)=1, K=1, state $s=(\{\delta_1^{E_{1}^{1}}\}, \{x_1^{1}\}, e)$).}
%\label{fig:transition}
%\vspace{-0.2 cm}
%\end{figure*}

\section{SMDP-based Channel Allocation Model}\label{sec:SMDP}
In this section, we develop an SMDP-based path allocation model to study the performance of a RIS system considering the unreliability of RIS devices.
We aim  to take optimal decisions at every decision epochs; arrival of new service request, departure of a service, failure of a RIS, failure of a meta-surface,  return of a failed meta-surface block, where our goal is to maximize the long-term expected system rewards. 
 The expected discounted reward is given based on the model in \cite{puterman2014markov} as follows:
\begin{equation}
\begin{split}
r(s,a)=&w(s,a)-c(s,a)\cdot E_{s}^{a}\lbrace \int_{0}^{\tau} e^{-\alpha t} dt\rbrace\\
=&w(s,a)-c(s,a)\cdot E_{s}^{a}\lbrace \frac{1-e^{-\alpha \tau}}{\alpha}\rbrace\\
=&w(s,a)- \frac{c(s,a) }{\alpha + \tau(s,a)}
\end{split}
\end{equation}
where $\alpha$ is a continuous-time discount factor.

Using the defined transition probabilities eq. (\ref{eq:tr1}), (\ref{eq:tr2}), (\ref{eq:tr3}),  (\ref{eq:tr6}),(\ref{eq:tr8}), we can obtain
the maximum long-term discounted reward using a discounted reward model defined in \cite{puterman2014markov} as
\begin{equation}
\nu(s)=\max_{a\in A(s)}\left\lbrace r(s,a) + \lambda \sum_{s'\in S} p(s'|s,a) \nu(s') \right\rbrace
\end{equation}
where $\lambda=\tau(s,a) / (\alpha+\tau(s,a)) $.
In the SMDP model, the value of $\nu (s)$ in a  strategy $\psi$ is computed based on the value $\nu (s')$ obtained in the strategy $\psi - 1$, and as an initial value, the discounted reward can be set to zero for all states to initialize the computation, which will converge afterwards to the optimal solution.\\ 
To simplify the computation of the reward, let $\rho$ be a finite constant, where $\rho= \lambda_s  +  \lambda^{m}\prod_{i=1}^{R}N(i) +  \mu^{m}\prod_{i=1}^{R}N(i) + \mu_s \sum_{i=1}^{R} N(i)  < \infty$. We define $\overline{p}(s'|s,a)$, $ \overline{\nu}(s)$, and $\overline{r}(s,a)$  as the uniformed transition probability, long-term reward, and reward function, respectively, and given by:
\begin{equation}
\overline{r}(s,a)= r(s,a)\frac{\tau (s,a) + \alpha}{\rho+\alpha}, \overline{\lambda}= \frac{\rho}{\rho+\alpha}
\end{equation}
\begin{equation}
\begin{split}
\overline{p}(s'|s,a)= 
 \begin{cases}
       1-\frac{[1-p(s'|s,a)] \tau (s,a)}{\rho}  & s'=s\\
       \frac{p(s'|s,a) \tau (s,a)}{\rho} & s'\neq s\\
    \end{cases}
\end{split}
\end{equation}
 After uniformization, the optimal reward is given by:
\begin{equation}\label{eq:reward}
\overline{\nu}(s)=\max_{a\in A(s)}\left\lbrace \overline{r}(s,a) + \overline{\lambda} \sum_{s'\in S} \overline{p}(s'|s,a) \overline{\nu}(s') \right\rbrace
\end{equation}

In order to solve our SMDP-based path Allocation (PA) model, we consider an iterative algorithm described as follows:  
\begin{algorithm}
\caption{Iterative SMDP-PA Algorithm}
\label{alg:SMDPECalgorithm}
\begin{algorithmic}[1]
\State \textbf{Step 1 (Initialization):} $\overline{\nu}^0(s)=0$, for all $s\in S$. Set the value of  $\epsilon >0$, and iteration $t=0$.

\State \textbf{Step 2:} Using eq. \ref{eq:reward}, compute the discounted reward for each state $s$:
$$\overline{\nu}^{t+1}(s)=\max_{a\in A(s)}\left\lbrace \overline{r}(s,a) + \overline{\lambda} \sum_{s'\in S} \overline{p}(s'|s,a) \overline{\nu}^{t}(s') \right\rbrace$$
\State \textbf{Step 3:}\If{$\| \overline{\nu}^{t+1} - \overline{\nu}^{t} \|> \epsilon$ } $t\longleftarrow t+1$,  go to \textbf{Step 2}
\Else  \; go to \textbf{Step 4} 
\EndIf
\State \textbf{Step 4:} Compute the optimal policy for all $s\in S$  
$$d_{\epsilon}^{*}(s)\in \text{arg}\max_{a\in A(s)}\left\lbrace \overline{r}(s,a) + \overline{\lambda} \sum_{s'\in S} \overline{p}(s'|s,a) \overline{\nu}^{t+1}(s') \right\rbrace$$

\end{algorithmic}
\end{algorithm}

After obtaining the optimal policy from Algorithm \ref{alg:SMDPECalgorithm}, the steady states probabilities are computed using the following system of equations:
\begin{equation}
\begin{split}
\pi (P-J)=0,  \sum_{s\in S}\pi(s)=1
\end{split}
\end{equation}
where $\pi(s)$ represents the steady state probability at state $s$, $P$ is the transition probabilities matrix, considering the optimal policy $d_{\epsilon}^{*}$, and $J$ denotes the all-ones matrix.

\section{Numerical results}\label{sec:results}

In this section, we validate and evaluate the proposed SMDP-based unreliability-aware path allocation for VR over THz RIS system using a Python simulator to implement the model and the algorithms proposed. The parameters used for simulation are summarized as follows: $R=1-3$, $K=1-2$, $c=1$, $\epsilon=100$, $Q=150$, $Z=100$,$\alpha=0.1$, $\lambda_s=1-10$, $\mu_s=5$. %in Table \ref{tab:param}.  
To investigate the performance of the RIS system under different system settings, the simulation results present a function of the service arrival rate, service departure rate, RIS blocks failure rate, RIS blocks return rate.  We define 4 scenarios with following settings: scenario 1) the RIS network is operating using a two RISs: a main RIS with a single backup RIS ($R=1$) containing $5$ meta-surfaces ($N(1)=5$), and allowing users to allocate only one block for each VR service ($K=1$); scenario 2) the RIS network is operating using  two main RIS with two backup RISs ($R=2$), similar to scenario 1, $K=1$; scenario 3) the RIS network is operating with three main RISs and  three backup RISs  ($R=3$), the system allows the allocation of two blocks at maximum for a single VR user request ($K=2$), where each RIS contains 5 blocks ($N(r_i)=5, \forall i \in [1,2,3]$); and scenario 4) the RIS network is operating with three main RISs and  three backup RISs  ($R=3$), the system allows the allocation of two blocks at maximum for a single VR user request ($K=2$), whereby the amount of blocks in RISs was set to $N(r_1)=4, N(r_2)=3, N(r_3)=2$. Similar to \cite{li2017smdp}, the discount factor $\alpha$ is 0.1.

%\begin{table}[h!]
%  \begin{center}
%    \caption{Simulation Parameters in RIS System.}
%    \label{tab:param}
%    \begin{tabular}{l|c|c|r} % <-- Alignments: 1st column left, 2nd middle and 3rd right, with vertical lines in between
%    \hline
%    \textbf{Parameter} & \textbf{Value}   & \textbf{Parameter} & \textbf{Value}\\
%      \hline
%       $R$ & 1-3 & $\epsilon$ & 100\\
%      $K$ & 1-2 & $Q$ & $150$\\
%       $c$ & 1 & $Z$ & $100$\\
%      $\lambda_s$ & 1-10 & $\mu_s$ & 5  \\
%      $\lambda^{r}$ & 1 & $\mu^{r}$ & 0.01\\
%       $\alpha$ & 0.1  &  &\\
%      
%    \end{tabular}
%  \end{center}
%\end{table}
\begin{figure*}
  \centering
  \subfloat[Action probabilities]{\includegraphics[width=0.33\textwidth]{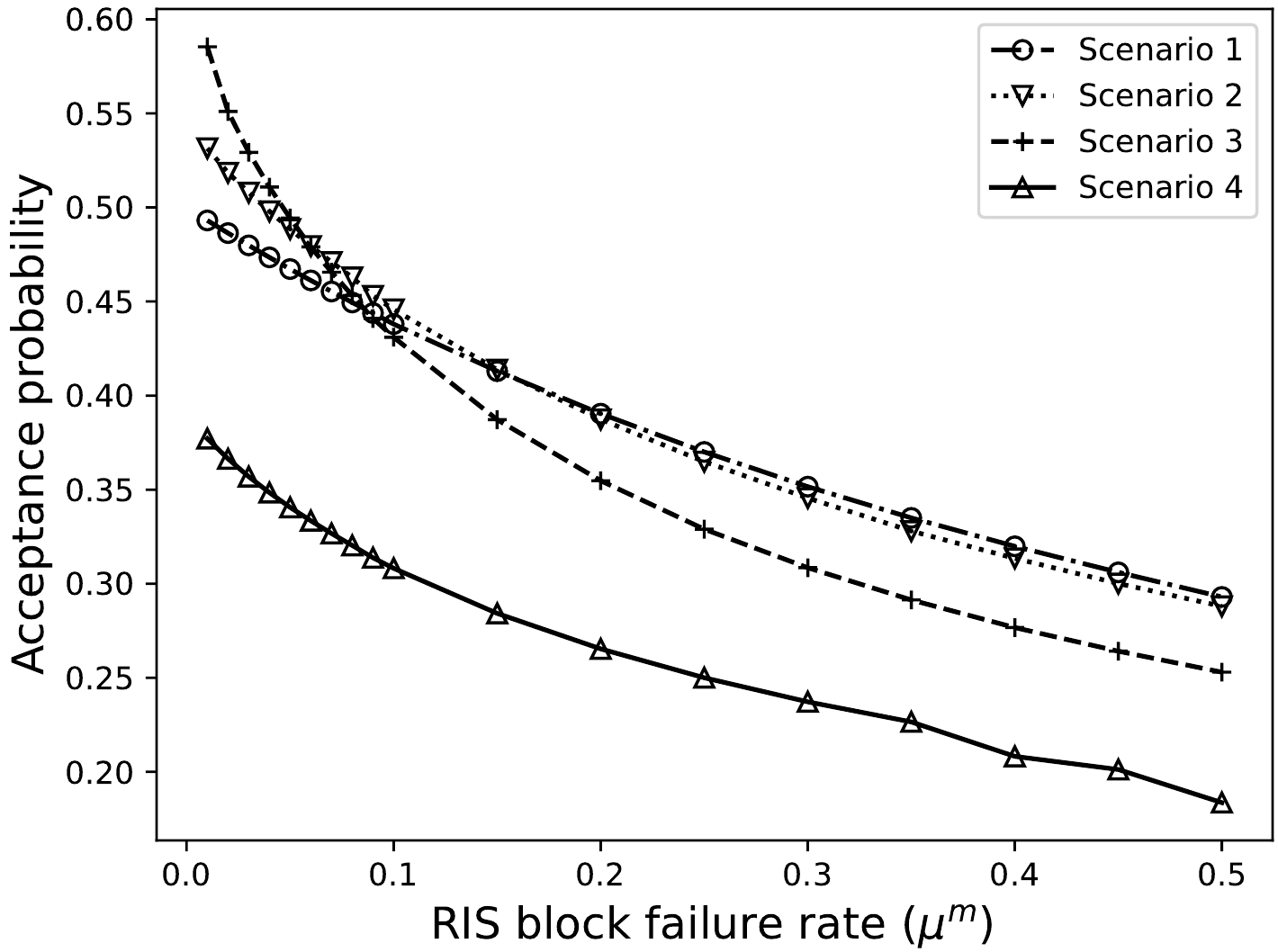}}
  \subfloat[Average system reward]{\includegraphics[width=0.33\textwidth]{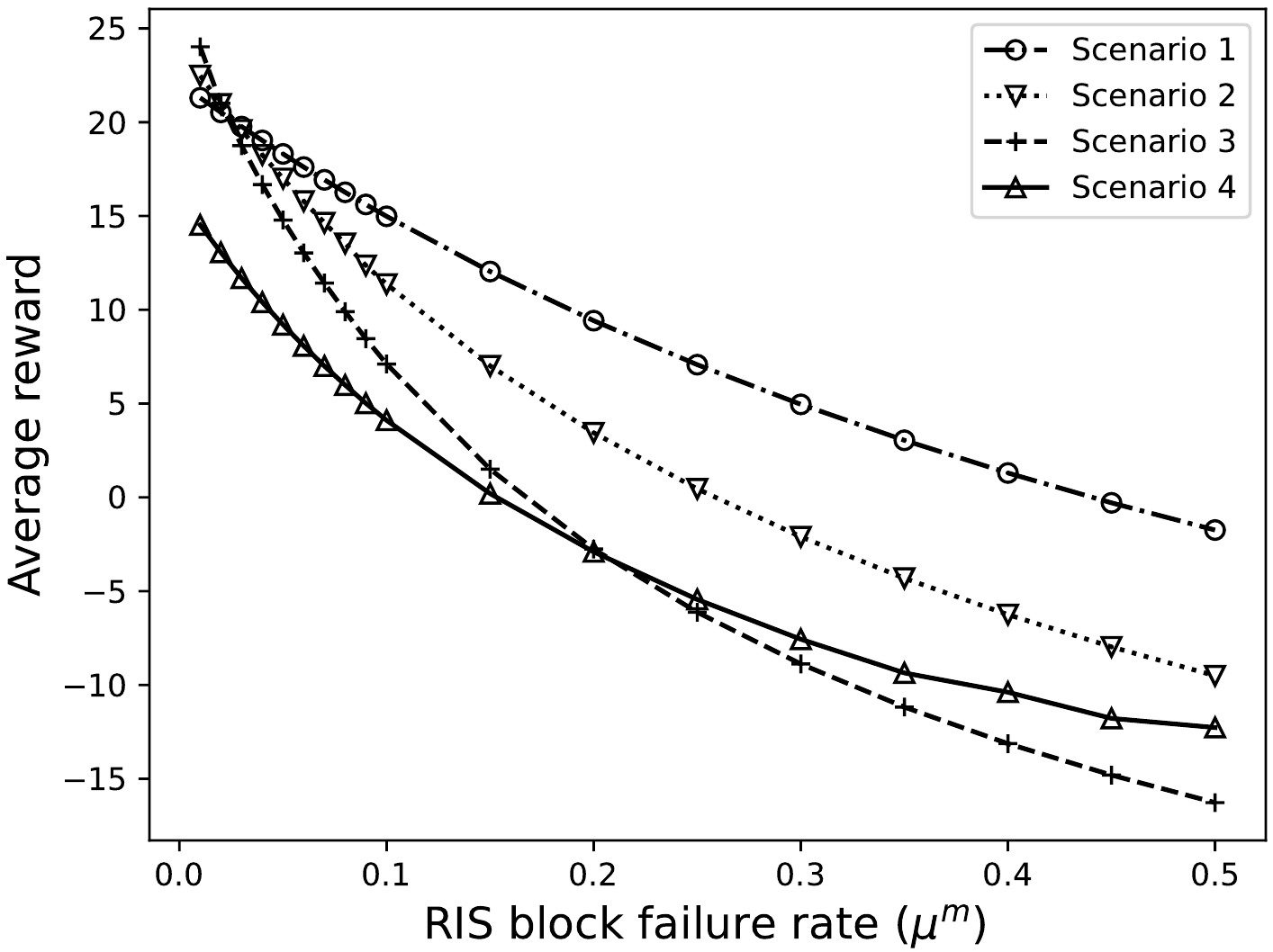}} 
  \subfloat[Blocking probabilities]{\includegraphics[width=0.33\textwidth]{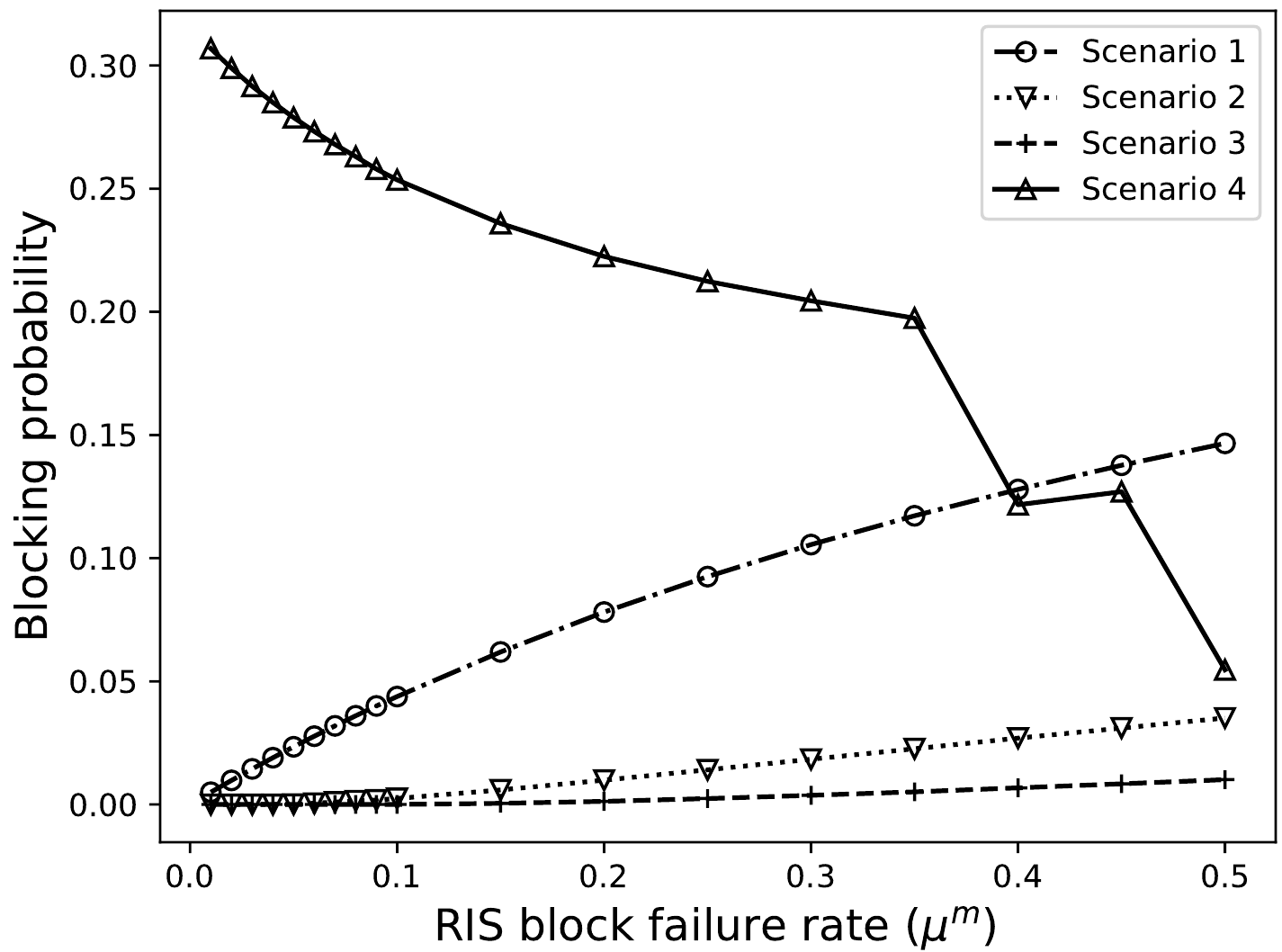}} 
  \caption{Results with  different RIS failure rates($\lambda_s=1$)}
 
\label{fig:res_fail}
\vspace{-0.3 cm}
\end{figure*}

%%%%%%%%%%%%%%%%%%%%%%%%%%%%%%%%%%%%%%%%%%%%%
Fig. \ref{fig:res_fail}.a) shows the acceptance probability as a function of the RIS block failure rate under different scenarios defined above. Here, we set $\lambda_s=1$. As expected, the acceptance probability of a service decreases with increasing failure rate of RIS blocks. The decrease of the acceptance probability can be explained with the increasing request blocking due to the capacity of the RIS system reduced by failed RIS blocks. The highest acceptance probability of around $68 \%$ could be reached in scenario 3 with 3 RISs and low RIS block failure rate. This can be explained by the fact that the high RIS blocks/paths availability with 4 RISs and 4 backup RISs, where hardware failure is managed by the backup RISs. The RIS system configured based on the scenario 1 and 2 show the highest acceptance probability for the higher RIS blocks failure rate starting from $\mu^{m}=0.1$ and decreases from ~45 \% to ~30\%. The RIS system in scenario 4 shows the lowest service acceptance probability up to ~37 \%, as the RIS system can not always satisfy the amount of VR user requests with the limited amount of RISs and  additional RIS failures.

%\begin{figure}
% \centering
%   \includegraphics[width=0.35\textwidth]{figures/actionprob_RISfailure.pdf}
% \caption{Action probabilities with different RIS failure rates ($\lambda_s=1$).}
%\label{fig:ap-node-failure}
%\vspace{-0.2 cm}
%\end{figure}

\begin{figure}
 \centering
   \includegraphics[width=0.35\textwidth]{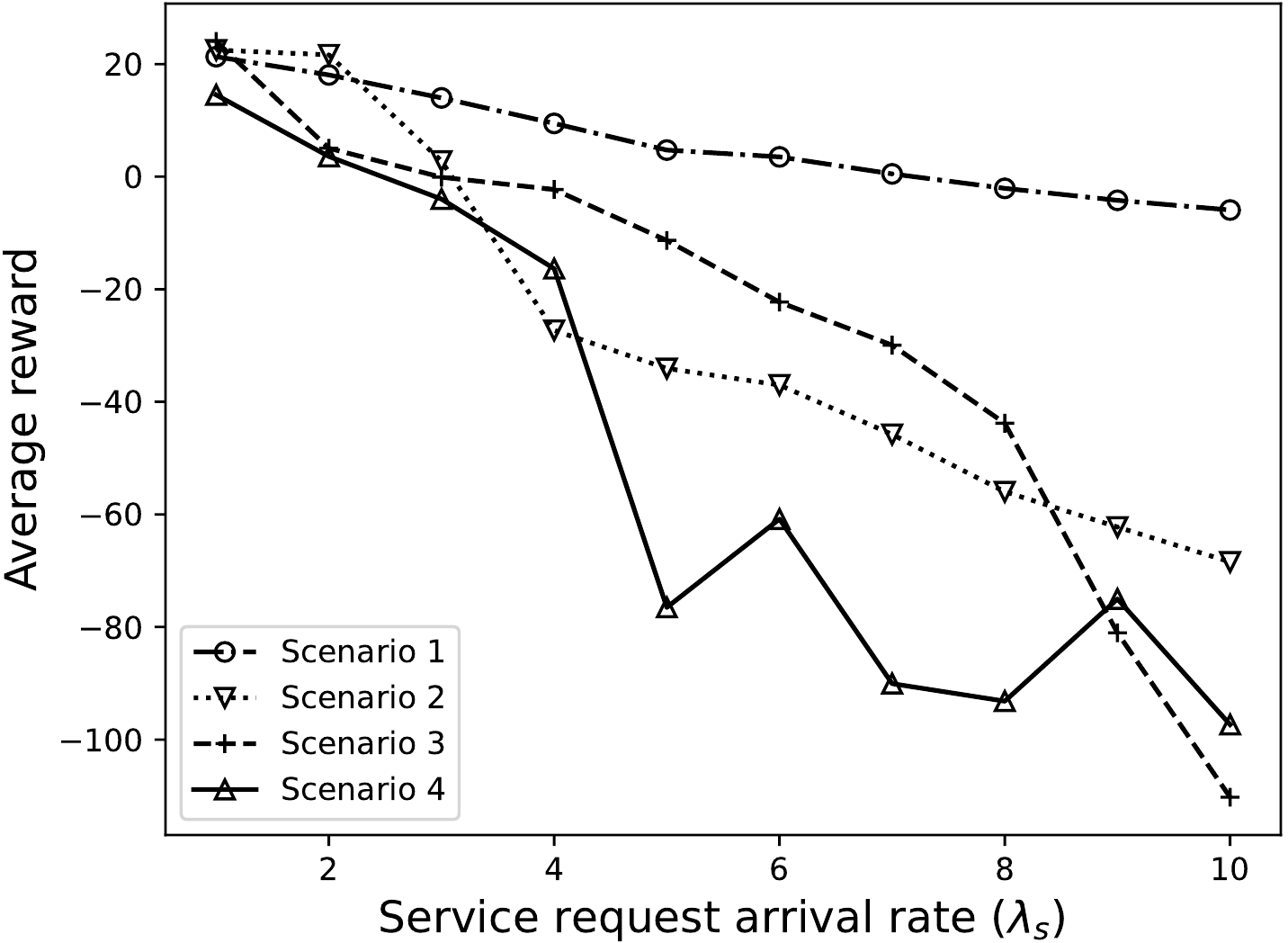}
 \caption{Average system reward with different arrival rates of services.}
\label{fig:avgr-service-arrival}
\vspace{-0.2 cm}
\end{figure}

Next we investigate the average system reward as a function of different arrival rates of service requests and different RIS block failure rates. Fig. \ref{fig:avgr-service-arrival} illustrates that the highest and the lowest average reward could be provided in scenario 1, and in scenarios 4 and 3 with $\lambda_s>9$, respectively. In general, the average reward decreases with increasing arrival rate of service requests, which is a result of the limited  capacity of RIS system. Additionally, in scenarios 3 and 4, the service requests can be blocked, when available RIS do not provide enough meta-surfaces/channels for the requested service. When the service arrival rate increases, the overall capacity needed to provide all service requests is higher than the RIS capacity. As a result, the system rejects any new incoming service requests, which decreases the average reward value. The average reward is a good metric for future work to compare an optimal path allocation policy with heuristic or machine learning based allocation algorithms.

Fig. \ref{fig:res_fail}.b) shows an average reward as a function of the RIS block failure rate. In our settings we highly penalize the RIS block failure, which explains the drastic decrease of the average reward, since the probability of accepting services becomes lower, and where we increased the number of RIS.  In Fig. \ref{fig:res_fail}.b) the average reward follows a convex decreasing function in terms of RIS block failure rates, with service arrival rate $\lambda_s$ equal to $1$. The figure shows that with low failure rate the reward is the same under all scenarios, and then it becomes the worst with scenarios that have more RIS. The minimum reward for scenario 1 is 0, which means that the single RIS  has failed and no service is accepted, while for scenario 3, the system can have failed RIS and still accepts service which will have a high probability to be lost in the case of failure, causing a negative reward value.

%%%%%%%%%%%%%%%%%%%%%%%%%%%%%%%%%%%%%%%%%%%%%
%%%%%%%%%%%%%%%%%%%%%%%%%%%%%%%%%
\begin{figure}
 \centering
   \includegraphics[width=0.35\textwidth]{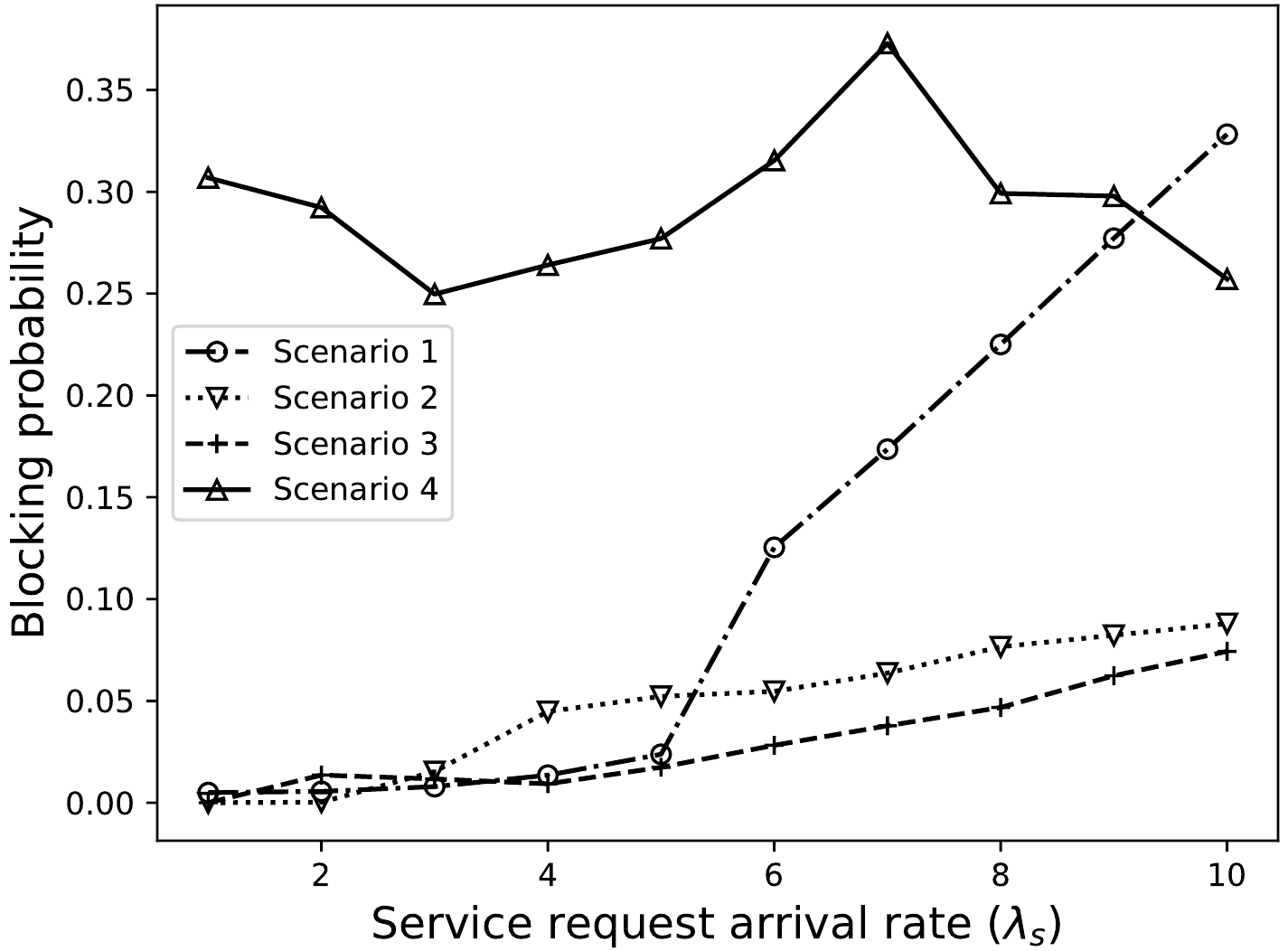}
 \caption{Blocking probabilities with different arrival rates of services.}
\label{fig:b-service-arrival}
\vspace{-0.2 cm}
\end{figure}
%\begin{figure}
% \centering
%   \includegraphics[width=0.35\textwidth]{figures/reward_RISfailure.pdf}
% \caption{Average system reward with different RIS failure rates($\lambda_s=1$).}
%\label{fig:avgr-node-failure}
%\vspace{-0.2 cm}
%\end{figure}

%%%%%%%%%%%%%%%%%%%%%%%%%%%%%%%%%%%%%%%%%%%%%
Finally we illustrate in Fig. \ref{fig:b-service-arrival}-\ref{fig:res_fail}.c) the  blocking probabilities with different arrival rates of services and with different RIS failure rates under different scenarios. We vary the service arrival rate and $\lambda_s=1$ when varying the RIS block failure rate. The blocking probability is very low with low arrival rates, less than $0.001$. It increases sharply after reaching the value $\lambda_s=5$, in scenario 1, due to the lack of blocks to allocate of a single RIS system and reaches the value $0.31$ when $\lambda_s=10$. For high service arrival rate, the more we increase the number of RISs the more stable our system gets. However, for low and medium arrival rates, scenario 1 with less RISs performs better than scenario 4 with 3 RISs, thus increasing the number of RISs and meta-surface blocks does not always improve the performance in terms of service blocking and long-term reward. In Fig. \ref{fig:res_fail}.c), we fix the service arrival rate to 1 and we vary the RIS blocks failure rate, we remark that the blocking probability is equal to 0 all values under $\mu^{m}=0.1$ which shows the stability of our channel allocation algorithm. For $\mu^{m}=0.5$, the blocking increases then to reach a maximum of $0.14$,  $0.03$ and $0.01$ for scenarios 1, 2 and 3, respectively. We see that increasing the number of RISs and meta-surface blocks highly improves the performance of RIS network in terms of service blocking.

%\begin{figure}
% \centering
%   \includegraphics[width=0.35\textwidth]{figures/RISfailure_blocking.pdf}
% \caption{Blocking probabilities with different RIS failure rates ($\lambda_s=1$).}
%\label{fig:b-node-failure}
%\vspace{-0.2 cm}
%\end{figure}
%%%%%%%%%%%%%%%%%%%%%%%%%%%%%%%%%%%%%%%%%%%%%%%%%%%%%%%%%%%%%%%%%%%%%%%%%%%%%%%%%%%%%%%%%%

 \section{Conclusion}\label{sec:conclusion}
We considered path allocation in the virtual reality (VR) applications over RIS  network with controlled access of VR users request by Semi-Markov decision Process (SMDP). We introduced an optimal path allocation scheme to ensure the reliability and maximize the system reward in a set of RISs, in which a RIS device and its meta-surface elements are vulnerable to failures. We formulated the problem as an SMDP model considering multiple RIS devices in an indoor environment used to provide services to VR users. Numerical results showed the average reward and blocking probability with various service arrival and RIS block failure rates, and under different network configurations.  The system computed corresponding policies to each service arrival rate to ensure the reliability and maximize long-term rewards. We showed that the proposed scheme was generally applicable, dynamic and provided an efficient solution to the path allocation problem. %Our proposed model works for online channel assignment to users using an iterative solution, that assumes the knowledge about some system statistics, such as the RIS failure,  service arrival, and service departure.  
%We showed that our scheme improved the QoS of VR users with regard to their arrival rate and RIS meta-surface blocks availability. Moreover, when VR user arrival rate were high, the channels were allocated with the minimum requirements for a VR service to operate, which decreased the blocking probability.  %\hl{what is the MAIN Finding from the results/ what is hte main avenue for future work???}
We showed that increasing the number of RISs and meta-surface blocks does not always improve the performance in terms of service blocking and long-term reward, for low and medium service arrival rate. Such finding allows us to better dimensioning RIS networks based on the expected service rate, in order to improve costs and performance at the same time. In our future work,  
a large-scale solution based on reinforcement learning can be proposed to solve value iteration algorithm, which is known by its exponential growth.% Also, the RIS system can consider VR users location to prioritize RIS associations before allocating channels. Finally, failure dependency between RIS devices can be considered in future system design.

\section*{Acknowledgment}
This work was partially supported by the DFG Project Nr. JU2757/12-1,  and by the Federal Ministry of Education and Research of Germany, joint project 6GRIC, 16KISK031.

\bibliographystyle{IEEEtran}
\bibliography{mybib}

\end{document}